\documentstyle[12pt,citesort,epsfig]{elsart}
\begin{document}
\begin{frontmatter}

\title{The Majorana neutrino masses, neutrinoless double beta decay 
and nuclear matrix elements}

\author[SISSA]{S.M. Bilenky}\footnote{On leave of absence from
the Joint Institute for Nuclear Research, 
141980 Dubna (Moscow Region), Russia
},
\author[Tuebingen]{Amand Faessler}, 
\author[Tuebingen]{F. \v Simkovic}\footnote{On  leave of absence from
Department of Nuclear Physics, Comenius University,
Mlynsk\'a dolina F1, SK--842 15 Bratislava, Slovakia}  

\address[SISSA]{SISSA, via Beirut 2-4, Trieste, 34014, Italy}

\address[Tuebingen]{Institut f\"ur Theoretische Physik, Universit\"at
T\"ubingen, Auf der Morgenstelle 14, D-72076 T\"ubingen, Germany}

\date{\today}
 
\maketitle
 
\vskip.5cm
 
\begin{abstract}
The effective Majorana neutrino mass $m_{\beta\beta}$ is 
evaluated by using the latest results of neutrino oscillation 
experiments. The problems of the neutrino mass spectrum, 
absolute mass scale of neutrinos and the effect of CP phases
are addressed. A connection to the next generation of the 
neutrinoless double beta decay ($0\nu\beta\beta$-decay)
experiments is discussed.
The calculations are performed for $^{76} \rm{Ge}$, $^{100}\rm{Mo}$, 
$^{136} \rm{Xe}$ and $^{130} \rm{Te}$ by using the advantage
of recently evaluated nuclear matrix elements with 
significantly reduced theoretical uncertainty. An importance 
of observation of the $0\nu\beta\beta$-decay of several nuclei 
is stressed. 

\vskip .3cm
 
 
\end{abstract}

\end{frontmatter}

\section{Introduction}

Strong evidence in favor of neutrino oscillations and small
neutrino masses were obtained in the Super-Kamiokande \cite {S-K}, SNO
\cite{SNO}, KamLAND \cite{Kamland} and other atmospheric 
\cite {Soudan,MACRO} and solar
\cite{Cl,GALLEX-GNO,SAGE,S-Ksol} neutrino 
experiments. The data of 
all these experiments are perfectly
described by the three-neutrino mixing 
\footnote{There exist at present indications in favor of 
$\bar\nu_{\mu} \to \bar\nu_{e}$ transitions, obtained 
in the accelerator LSND experiment \cite{LSND}. 
The LSND data can be explained by neutrino oscillations with
$\Delta m^{2}_{\rm{LSND}}\simeq 1\,~ \rm{eV}^{2}.$
The result of the LSND experiment will be checked by the MiniBooNE 
experiment at Fermilab \cite{MiniB}.}
\begin{equation}
\nu_{lL} = \sum_{i=1}^{3} U_{li} \nu_{iL};\,~l=e,\mu,\tau,  
\label{eq:1}
\end{equation}
where $\nu_{i}$ is the field of neutrino with mass $m_i$ and 
$U_{li}$ are the elements of the 
Pontecorvo--Maki--Nakagawa--Sakata (PMNS) unitary neutrino 
matrix \cite{ponte}. From the global analysis of the 
solar and KamLAND data \cite{SNOsalt} and Super-Kamiokande
atmospheric data \cite{S-K} 
the following best-fit values of the two independent neutrino
mass-squared differences were obtained
\begin{equation}
\Delta m^{2}_{\rm{sol}}= 7.1\,~10^{-5}\,~ \rm{eV}^{2},\,~
\Delta m^{2}_{\rm{atm}}= 2.0\,~10^{-3}\,~ \rm{eV}^{2}.
\label{eq:2}
\end{equation}

The observation of neutrino oscillations means
that the flavor lepton numbers $L_{e}$, $L_{\mu}$ and $L_{\tau}$
are not conserved by the neutrino mass term. If the total lepton number 
$L = L_{e} + L_{\mu} +L_{\tau}$ is conserved, 
neutrinos with definite masses $\nu_{i}$ are Dirac
particles.
If there are no conserved lepton numbers, $\nu_{i}$ 
are Majorana particles. 
The problem of the nature of massive neutrinos (Dirac or Majorana?) 
is one of the most fundamental one. The solution of this problem 
will have very important impact on the understanding of the origin 
of neutrino masses and mixing.

The investigation of the flavor neutrino oscillations 
$\nu_{l}\to \nu_{l'}$ does not allow to reveal the nature
of massive neutrinos $\nu_{i}$ \cite{BHP,valle1}. 
This is possible only via the investigation of the processes in which 
the total lepton number $L$ is not conserved. The
neutrinoless double $\beta$-decay  
\cite{doi,v5,suho98,v6,vog02},
\begin{equation}
(\rm{A,Z}) \to (\rm{A,Z +2}) +e^- +e^-
\label{eq:3}
\end{equation}
is the most sensitive process to the violation of the total 
lepton number and small Majorana neutrino masses.

By assuming the dominance of the light neutrino mixing 
mechanism\footnote{Note that, there are many other 
$0\nu\beta\beta$-decay mechanisms triggered by exchange of heavy neutrinos, 
neutralinos, gluinos, leptoquarks etc. \cite{doi,v5,moha1,moha2,SUSY,SUSY2,lepto}.
However, the observation of the 
$0\nu\beta\beta$-decay would mean that neutrinos are massive 
Majorana particles irrespective what is the mechanism of this process
\cite{valle2}.}
the inverse value of the $0\nu\beta\beta$-decay half-life 
for a given isotope (A,Z) is given by \cite{doi,v5,suho98,v6,vog02}
\begin{equation}
\frac{1}{T^{0\nu}_{1/2}(A,Z)}=
|m_{\beta \beta}|^{2}\,|M^{0\,\nu}(A,Z)|^{2}\, g_A^4\, 
G^{0\,\nu}_{01}(E_{0},Z)\,.
\label{eq:4}
\end{equation}
Here, $G^{0\,\nu}(E_{0},Z)$, $g_A$ and $|M^{0\,\nu}(A,Z)|$ are, 
respectively, the known phase-space factor ($E_{0}$ is the energy release),
the effective axial-vector coupling constant and the 
nuclear matrix element, which depends on the nuclear
structure of the particular isotope under study.
The main aim of the experiments on the search for
$0\nu\beta\beta$-decay is the measurement of the effective 
Majorana neutrino mass  $m_{\beta \beta}$.

Under the assumption of the mixing of three massive Majorana neutrinos 
the effective Majorana neutrino mass 
$m_{\beta \beta}$ takes the form 
\begin{equation}
m_{\beta\beta} = U^{2}_{e1}\,m_{1} + U^{2}_{e2}\,m_{2} +U^{2}_{e3}\,m_{3}.
\label{eq:5}
\end{equation}
The predictions for $m_{\beta \beta}$ can be obtained
by using the present data on the oscillation parameters.
Its value depends strongly on the type of neutrino mass spectrum 
and minimal neutrino mass \cite{hier2,hier1,hier3,crit,hiera,hier5,hier6,hier4,hier7,hier8}.

The $0\nu\beta\beta$-decay has not been seen experimentally till now. 
The best result have been achieved in the Heidelberg--Moscow
(HM) $^{76}Ge$ experiment \cite{HM} ($T_{1/2}^{0\nu} \ge 1.9 ~ 10^{25}$ years).
By assuming the  $0\nu\beta\beta$-decay matrix element
of Ref. \cite{rodin} and the result of the HM experiment \cite{HM} 
we end up  with upper limit on the effective 
Majorana mass $|m_{\beta\beta}| \le 0.55~eV$. 
Recently, some  authors of the HM collaboration have claimed  
the experimental observation  of the $0\nu\beta\beta$-decay of
$^{76}Ge$ with half-lifetime $ T^{0\nu}_{1/2} = (0.8-18.3)~ 10^{25}$ years 
(best-fit value of $1.5~ 10^{25}$ years) 
\cite{evid}\footnote{
Note that, the Moscow participants of the HM
collaboration performed a separate analysis of 
the data and presented the results  at NANP 2003 
(Dubna, June 24, 2003) \cite{nanp}. They found no indication in 
favor of the evidence of the $0\nu\beta\beta$-decay.}. 
This work has attracted a lot of attention of both experimentalists 
and theoreticians due to 
important consequences for particle physics and astrophysics \cite{sark}. 
Several researchers of the $\beta\beta$- 
decay community re-examined and critized the 
paper, suggesting a definitely weaker statistical significance 
of the peak \cite{crit,yzd,aal}. 
In any case the disproof or the confirmation of the claim 
will come from future experiments. 
A good candidate for a cross-check of the claimed evidence of the
$0\nu\beta\beta$-decay of $^{76}Ge$ is the Cuoricino/CUORE experiment \cite{cuorefu}
in which $0\nu\beta\beta$-decay of $^{130}Te$ is investigated.  

There are many  other ambitious projects in preparation,
in particular  CAMEO, CUORE, COBRA, EXO, GEM, GENIUS, MAJORANA, 
MOON, XMASS etc \cite{vog02,cuorefu,zdefu,klfu}. In the   next generation 
$0\nu\beta\beta$-decay detectors a few tons of 
the radioactive $0\nu\beta\beta$-decay 
material will be used.  
This is a very big improvement as the current experiments use only 
few tens of kg's for the source. 
The future double beta decay experiments stand to uncover 
the fundamental nature of neutrinos (Dirac or Majorana), probe the mass 
pattern, and perhaps determine the absolute neutrino mass scale 
and look for possible CP violation. 

The uncertainty in $m_{\beta \beta}$ is an important issue. The precision 
of the oscillation parameters is expected to be significantly improved
in the future neutrino experiments  at the JPARC facility \cite{koba}
in new reactor neutrino experiments \cite{reactor}
in off-axes neutrino experiments \cite{feld}
in the $\beta$-beam experiments  \cite{beam}
and  Neutrino Factory experiments \cite{fact}.
The primary concern are the nuclear matrix elements.
Clearly, the accuracy of the determination of the effective Majorana mass from the
measured $0\nu\beta\beta$-decay 
half-life is mainly determined by our the knowledge of the nuclear matrix elements. 
Reliable nuclear matrix elements are required as they guide future choices
of isotopes for the $0\nu\beta\beta$-decay experiments. 

In this article the problem of the uncertainty of the $0\nu\beta\beta$-decay
matrix elements will be addressed. A further development in the calculation
of the  $0\nu\beta\beta$-decay ground state transitions will be indicated.
By using the latest values of the neutrino oscillation parameters
the possible values of the effective Majorana mass $|m_{\beta\beta}|$
will be calculated. By using the nuclear matrix elements of Ref. \cite{rodin}
with reduced theoretical uncertainty, the perspectives
of the proposed  $0\nu \beta\beta$-decay experiments 
(CUORE, GEM, GENIUS,Majorana, 
MOON, EXO and XMASS) in discerning 
the normal, inverted and almost degenerate 
neutrino mass spectra will be studied.

The paper is organized as follows:
In Section 2 the problem of the calculation of the 
$0\nu\beta\beta$-decay matrix elements will be discussed. 
The sensitivity of future $0\nu\beta\beta$-decay experiments
to the lepton number violating parameter $m_{\beta \beta}$ 
will be established. In Section 3 the effective Majorana neutrino
mass $m_{\beta \beta}$ will be calculated by using the data
of neutrino oscillation experiments and assumptions about character 
of neutrino mass spectrum. Conclusions on the 
discovery potential of planned $0\nu\beta\beta$-decay experiments
will be drawn. In Section 3 we present the summary and our
final conclusions.

\section{Uncertainties of the $0\nu\beta\beta$-decay nuclear matrix elements}

The reliable value (limit) for the 
fundamental particle physics 
quantity $m_{\beta \beta}$ can be inferred from experimental data only if 
 the nuclear matrix elements governing 
the $0\nu\beta\beta$-decay are calculated correctly, i.e., 
the mechanism of nuclear transitions is well understood \cite{v5,suho98}. 

The nuclear matrix element $M^{0\,\nu}(A,Z)$ is  given as a sum
of Fermi, Gamow-Teller, and  tensor contributions:
\begin{equation}
M^{0\nu}(A,Z) = -\frac{M^{0\nu}_F(A,Z)}{g^2_A} + 
M^{0\nu}_{GT}(A,Z) + M^{0\nu}_T(A,Z).
\label{eq:6}   
\end{equation}
The explicit form of the particular matrix elements ${M^{0\nu}_F}$,
$M^{0\nu}_{GT}$ and $M^{0\nu}_T$ can be found in Ref. \cite{simk}. 
In this work in comparison to the most of $0\nu\beta\beta$-decay 
studies \cite{VZ86,CFT87,MK88,found,suho,shellm}
the higher order terms of the nucleon current were 
taken into account. Their contributions result in suppression
of the nuclear matrix element $M^{0\nu}$
by about 30 \% for all nuclei. The weak axial coupling 
constant $g_A$, which reduces the $M^{0\nu}_{F}$ contribution 
to the $0\nu\beta\beta$-decay matrix element, 
is one of sources of the uncertainty in the determination
of $M^{0\nu}$. Usually, it is fixed at $g_A = 1.25$
but a quenched value $g_A = 1.0$ is also considered. 
The estimated uncertainty of $M^{0\nu}$
due to $g_A$ is of the order of 20 \% \cite{simk}.

The evaluation of the nuclear matrix element $M^{0\nu}$
is a complex task. The reasons are:\\ 
i) The nuclear systems which 
can undergo double beta decay are medium and heavy open-shell nuclei with 
a complicated nuclear structure. One is forced to introduce 
many-body approximations in solving this problem.\\
ii) The construction
of the complete set of the states of the intermediate nucleus is needed as 
the $0\nu \beta\beta$-decay  is a second order in the weak interaction process.\\ 
iii) The confidence level of the nuclear structure parameter choice
have to be determined. There are many parameters 
entering the calculation of nuclear matrix elements, in particular
the mean field parameters, pairing interactions, 
particle-particle and particle-hole strengths,
the size the model space, nuclear deformations etc. 
It is required to fix them
by a study of associated nuclear processes like single $\beta$ and
$2\nu\beta\beta$-decay, ordinary muon capture and others. This 
procedure allows to assign the level of significance to
the calculated $0\nu\beta\beta$-decay matrix elements.

The nuclear wave functions can be tested, e.g., by calculating 
the two neutrino double beta decay ($2\nu\beta\beta$-decay) 
\begin{equation}
 (A,Z) \rightarrow   (A,Z+2) + 2e^-  + 2{\tilde{\nu}}_e,
\label{eq:7}
\end{equation}
for which we have experimental data. The $2\nu\beta\beta$-decay has been
directly observed so far in ten nuclides and into one excited state
\cite{data}.  
The inverse half-life of the $2\nu\beta\beta$-decay can be expressed as
a product of an accurately known phase-space factor $G^{2\,\nu}_{01}(E_{0},Z)$
and the Gamow-Teller transition matrix element $M^{2\,\nu}_{GT}(A,Z)$, 
which is a quantity of the second order in the perturbation theory:
\begin{equation}
\frac{1}{T^{2\nu}_{1/2}(A,Z)} =
|M^{2\,\nu}_{GT}(A,Z)|^{2}\, g_A^4\, G^{2\,\nu}_{01}(E_{0},Z).
\label{eq:8}
\end{equation}
The contribution from two successive Fermi transitions is safely neglected
as they come from isospin mixing effect. As $G^{2\,\nu}_{01}(E_{0},Z)$
is free of unknown parameters, the absolute value of 
the nuclear matrix element $M^{2\,\nu-exp}_{GT}(A,Z, g_A)$ can be extracted 
from the measured $2\nu\beta\beta$-decay half-life for a given $g_A$.

There are two well established approaches for the
calculation of the double beta decay nuclear matrix elements, namely 
the shell model \cite{shellm}
and the Quasiparticle Random Phase Approximation (QRPA) \cite{v5,suho98}.
The two methods differ in the size of model spaces and the way
the ground state correlations are taken into account. 
The shell model describes only a small energy window of the 
lowest states of the intermediate nucleus, but in a precise way. 
The significant truncation of the model space does not allow 
to take into account the $\beta$ strength from the region of the Gamow-Teller
resonance, which might play an important role.
Due to a finite model space one is forced to introduce effective 
operators, a procedure which is not well under control yet \cite{effop}.
During the period of the last eight years no progress in the shell model 
calculation of the double beta decay transitions were acknowledged. 

The QRPA plays a prominent role in the analysis, which is
unaccessible to shell model calculations. It is the most commonly
used method for calculation of double beta decay rates 
\cite{v5,suho98,VZ86,CFT87,MK88,found,suho}. The question is how
accurate is it? For a long period it was considered that 
predictive power of the QRPA approach is limited because of the
large variation of the relevant $\beta\beta$ matrix elements  
in the physical window of particle-particle strength of nuclear Hamiltonian. 
Many new extensions of the standard proton-neutron QRPA (pn-QRPA),
based on the quasiboson approximations, were proposed:\\
i) {\it The renormalized proton-neutron QRPA} (pn-RQRPA) \cite{toi95,pauli}. 
By implementing the Pauli exclusion principle (PEP)
in an approximate way in the pn-QRPA, one gets pn-RQRPA, which avoids
collapse within a physical range of the particle-particle force and offers
a more stable solution. The price paid for it is a small violation
of the Ikeda sum rule (ISR) which seems to have only small impact on the
calculation of the double beta decay matrix elements.  The studies 
performed within the schematic proton-neutron Lipkin model 
\cite{ppff} and realistic calculations of 
$M^{2\,\nu}_{GT}$ \cite{v5} proved that the RQRPA 
is more reliable method than the pn-QRPA.\\
ii) {\it The QRPA with proton-neutron pairing \cite{cheoun}
and the full RQRPA \cite{found,pauli}}. The modification of the 
quasiparticle mean field due to proton-neutron (pn) pairing
interaction affects the single $\beta$ and the $\beta\beta$ transitions. 
There are some open questions concerning fixing of 
the strength of the proton-neutron pairing. Recently, it was confirmed within the
deformed BCS approach that for nuclei with N much bigger than Z 
the effect of proton-neutron pairing is small but not negligible
\cite{pnp}. There is a possibility to consider simultaneously both the
pn pairing and the PEP within the QRPA theory. This version of the  QRPA 
is denoted as the full RQRPA \cite{pauli} in the  literature. \\
iii) {\it The proton-neutron self-consistent RQRPA (pn-SRQRPA) \cite{lublin}}. 
The pn-SRQRPA goes a step further beyond the pn-RQRPA by at the same
time minimizing the energy and fixing the number of particles 
in the correlated ground state instead of the uncorrelated BCS one
as is done in other versions of the QRPA. However, a large effect found 
in the $\beta\beta$-transitions with realistic NN-interaction \cite{lublin}
is apparently associated with consideration of bare pairing forces,
not fitted to the atomic mass differences,
within a complicated numerical procedure \cite{frq}.\\ 
iv) {\it The deformed QRPA}. Almost all  current 
$\beta\beta$-decay calculations for  nuclei of
experimental interest were performed by assuming spherical symmetry. 
Recently, an effect of deformation on the $2\nu\beta\beta$-decay 
matrix elements were studied within the deformed QRPA.
A new suppression mechanism of the $2\nu\beta\beta$-decay  matrix elements 
based on the difference in deformations of the initial and final nuclei 
were found  \cite{dqr}. It is expected that this effect might be important 
also for the $0\nu\beta\beta$-decay transitions.\\

The QRPA many-body approach for description of nuclear transitions 
is under continues development. 
In particular, it has been found feasible to include non-linear terms in 
the phonon operator \cite{nlf}. An another  modification of the QRPA 
phonon operator, which allows to fulfill the ISR exactly, were proposed 
in Ref. \cite{vaam}. Thus, a further progress in the QRPA calculation
of the $\beta\beta$-decay matrix elements is expected.

To estimate the uncertainty of the $0\nu\beta\beta$-decay transition
probability, different groups performed calculations in the
framework of different 
methods (pn-QRPA, pn-RQRPA, pn-SRQRPA, FRQRPA, QRPA with pn pairing
and deformed QRPA), different model spaces and different realistic forces. 
One might obtain  in this way an uncertainty by a factor 2 to 3, depending
especially on the method and the size of the model space. 
However, a significant progress has been achieved in the calculation of the
$0\nu\beta\beta$-decay matrix elements \cite{rodin} recently. 
It was shown that by fixing
the strength of the particle-particle interaction, so that the measured
$2\nu\beta\beta$-decay half-life is correctly reproduced, the resulting 
$M^{0\nu}$ become essentially independent on the
considered NN-potential, size of the basis and the restoration of the
PEP. The uncertainty of the obtained results for A=76, 100, 130 and 136 nuclei 
has been found less than 10  \%. This an exciting development. It is
desired to extend this type of study also to other nuclei 
and  other extensions of the QRPA approach. In this
way a correct  understanding of the uncertainty of the  
$0\nu\beta\beta$-decay matrix elements evaluated within the QRPA theory can be
established. 

The small spread of the $0\nu\beta\beta$-decay 
results obtained within the procedure 
of Ref. \cite{rodin} can be qualitatively understood.  It seems that
there is an advantage to consider the ratio of the $2\nu\beta\beta$-decay
and $0\nu\beta\beta$-decay matrix elements for a given isotope,
\begin{equation}
{\cal R}^{2\nu/0\nu}(A,Z) = | \frac{M^{2\,\nu}_{GT}(A,Z)}
{M^{0\nu}(A,Z)}|,
\label{eq:9}
\end{equation}
as in this quantity the dependence  on the some nuclear structure 
degrees of freedom is suppressed. By assuming
\begin{equation}
M^{2\,\nu}_{GT}(A,Z) = M^{2\,\nu-exp}_{GT}(A,Z,g_A)
\label{eq:10}
\end{equation}
the absolute value of the $0\nu\beta\beta$-decay matrix element 
can be inferred. We note that in comparison with $M^{2\,\nu}_{GT}$,
which is evaluated within a nuclear model, the value of $M^{2\,\nu-exp}_{GT}$, 
which is determined from the $2\nu\beta\beta$-decay half-life,
depends on $g_A$. The $2\nu\beta\beta$-decay plays a crucial
role in obtaining $0\nu\beta\beta$-decay matrix elements with a reduced
uncertainty  \cite{rodin}.

From the experimental upper limit on the $0\nu\beta\beta$-decay 
half-life $T^{0\nu-exp}_{1/2}(A,Z)$ it is straightforward 
to find a constraint on the effective Majorana neutrino mass 
$m_{\beta\beta}$ \cite{rodin}: 
\begin{equation}
|m_{\beta\beta}|~~ \le~~ 
[G^{0\,\nu}_{01}(E_{0},Z)~ T^{0\nu-exp}_{1/2}(A,Z)]^{-1/2}
~~ \frac{1}{g_A^2~ |M^{0\nu}(A,Z)|}.
\label{eq:11}
\end{equation}
In this work we consider the RQRPA $0\nu\beta\beta$-decay
matrix elements of Ref. \cite{rodin}, which were determined
with help of the  average values of the measured $2\nu\beta\beta$-decay
half-lives. They are given in the Table 1 of Ref. \cite{vog02}. 
In the case of $^{136}Xe$ for which the $2\nu\beta\beta$-decay
has been not observed yet, the current lower limit on the 
half-life is considered
as a reference. For the $^{130}Te$ isotope we took into account
recent measurement of the $2\nu\beta\beta$-decay half life
of $^{130}Te$ by the CUORE collaboration:
$T^{0\nu-exp}_{1/2} = (6.1 \pm 1.4~ (stat)~ + 2.9 - 3.5~ (sys)) ~ 10^{20}$ 
years \cite{cunew}.
This value  is by about a factor 3 smaller than
the previously considered average value given in Ref. \cite{vog02}. 
However, this has only a small
impact on the calculated $0\nu\beta\beta$-decay matrix element, 
which increases  by about 20 \%. 

In Table \ref{table.1} we present both the $0\nu\beta\beta$-decay 
\cite{rodin} and the $2\nu\beta\beta$-decay matrix elements, ratio 
${\cal R}^{2\nu/0\nu}(A,Z)$,
the average and measured $2\nu\beta\beta$-decay half-lives,
the current experimental limits  on the $0\nu\beta\beta$-decay 
half-life and the  half-lifetimes of the $0\nu\beta\beta$-decay,
which are expected  in future experiments after 5 years of data 
taking \cite{vog02}. By glancing at the Table \ref{table.1} we see that
the values of ${\cal R}^{2\nu/0\nu}$ for various
nuclei differ significantly each from other.
This is connected with the fact that the
$2\nu\beta\beta$-decay matrix element is  sensitive to the energy 
distribution of the $\beta$ strengths via the energy denominator. 
The largest value is associated with A=100 system for which the
ground state of the intermediate nucleus is the $1^+$ state. 

The current upper limits on the effective Majorana neutrino 
mass $m_{\beta\beta}$ and expected sensitivities of running
and planned experiments to this parameter 
for A=76, 100, 130 and 136 are listed in Table \ref{table.1}. 
We see that the Heidelberg-Moscow experiment  \cite{HM}  
offers the most restrictive limit  $|m_{\beta\beta}|  \le 0.55~eV$.
In future the sensitivity to  $|m_{\beta\beta}|$ might be
increased by about one order of magnitude (see Table 
\ref{table.1}).

If the $0\nu\beta\beta$-decay will be observed, the question of the
reliability of the deduced value on $|m_{\beta\beta}|$ will be 
a subject of great importance.
This problem can be solved by observation of the $0\nu\beta\beta$-decay
of several nuclei. Any uncertainty on the nuclear matrix element
reflects directly on measurement of  $|m_{\beta\beta}|$.
The spread of the $|m_{\beta\beta}|$ values associated with different
nuclei will allow to conclude about the accuracy
of the calculated $0\nu\beta\beta$-decay matrix elements.
An another scenario was proposed in Ref. \cite{nme}.  
It was suggested to study the ratios of $0\nu\beta\beta$-decay
matrix elements of different nuclei deduced from the corresponding 
half-lives. Unfortunately, in this way 
the uncertainty of the absolute value of the $0\nu\beta\beta$-decay
matrix elements can not be established. 
The first results obtained within the recently improved
QRPA procedure for calculating 
nuclear matrix element  \cite{rodin} are encouraging and
suggest that their uncertainty for a given isotope 
is of the order of tenths of percents. 
It comes without saying that it has to be confirmed by  further
theoretical analysis.

\section{The effective Majorana neutrino mass and neutrino oscillation data}

The effective Majorana mass is determined by the absolute values of neutrino
masses $m_{i}$ and elements of the first row of neutrino mixing matrix 
$U_{ei}$ (i=1,2,3). 
Taking into account existing 
neutrino oscillation data
we will discuss now a possible value of $|m_{\beta\beta}|$.

In the Majorana case all elements $U_{ek}$ are
complex
quantities 
\begin{equation}
U_{ek}=| U_{ek}| \,e^{i \alpha_{k} } ,                      
\label{eq:12}
\end{equation}
where $\alpha_{k}$ is the Majorana CP phase. If CP- invariance in the
lepton sector holds, we have
\begin{equation}
U_{ek}=U^*_{ek} \,\eta_{k},                      
\label{eq:13}
\end{equation}
where $\eta_{k}= i\,\rho_{k}\,~~(\rho_{k}=\pm 1)$ 
is the CP-parity of neutrino with definite mass .
Thus, in the case of the CP-invariance 
we have
\begin{equation}
2\,\alpha_{k}= \frac{\pi}{2}\,\rho_{k} .                     
\label{eq:14}
\end{equation}

Neutrino oscillation data are compatible with two types of neutrino
mass spectra\footnote{$\Delta m^{2}_{ik}$ is defined as follows: 
$\Delta m^{2}_{ik}=m^{2}_{i}-
m^{2}_{k}$}:
\begin{enumerate}
\item
"Normal" mass spectrum $m_{1}< m_{2}< m_{3}$

$$\Delta m^{2}_{21} \simeq\Delta m^{2}_{\rm{sol}};\hspace{1cm}
\Delta m^{2}_{32} \simeq\Delta m^{2}_{\rm{atm}}.$$

\item
"Inverted" mass spectrum $m_{3}< m_{1}< m_{2}$

$$\Delta m^{2}_{21} \simeq\Delta m^{2}_{\rm{sol}};\hspace{1cm}
\Delta m^{2}_{31} \simeq -\Delta m^{2}_{\rm{atm}}.$$
\end{enumerate}

For neutrino masses in the case of the normal spectrum we have
\begin{equation}
m_{2} \simeq \sqrt{m^{2}_{1}+\Delta m^{2}_{\rm{sol}}},~~~~
m_{3}  \simeq \sqrt{m^{2}_{1}+\Delta m^{2}_{\rm{atm}}},
\label{eq:15}
\end{equation}
where we took into account that $\Delta m^{2}_{\rm{sol}}\ll\Delta m^{2}_{\rm{atm}}$.
In the case of the inverted spectrum we have
\begin{equation}
m_{2} \simeq 
m_{1} \simeq \sqrt{m^{2}_{3}+\Delta m^{2}_{\rm{atm} }}.
\label{eq:16}
\end{equation}
The elements $|U_{ei}|^{2}$ for both types of neutrino mass spectra
are given by
\begin{equation}
|U_{e1}|^{2}=\cos^{2} \theta_{13}\, \cos^{2} \theta_{12},~~~
|U_{e2}|^{2}=\cos^{2} \theta_{13}\, \sin^{2} \theta_{12},~~~
|U_{e3}|^{2}=\sin^{2} \theta_{13}
\label{eq:17}
\end{equation}
The mixing angle $\theta_{12}$ was determined from the data of the
solar neutrino experiments and KamLAND reactor experiment. From the
latest analysis of the existing data for the best fit value of 
$\sin^{2}\theta_{12} $
it was found \cite{SNO}
\begin{equation}
\sin^{2} \theta_{12} \simeq \sin^{2} \theta_{\rm{sol}}= 0.29
\label{eq:18}
\end{equation}

For the angle $\theta_{13}$ only upper bound is known. From the
exclusion plot obtained from the data of the reactor experiment CHOOZ \cite{fogli}
at $\Delta m^2_{32} = 2\cdot 10^{-3}\,\rm{eV}^{2}$ (the Super-Kamiokande 
best-fit value) we have
\begin{equation}
\sin^{2} \theta_{13} \leq 5\cdot10^{-2}.
\label{eq:19}
\end{equation}
For the minimal neutrino mass $m_{1}$ ($m_{3}$) we know also
only an upper bound . From the data of the tritium  
Mainz \cite{mainz} and Troitsk \cite{troitsk} experiments 
it was found
\begin{equation}
m_{1} \leq 2.2 \,~\rm{eV}
\label{eq:20}
\end{equation}
In the future tritium experiment KATRIN \cite{katrin} the sensitivity
$m_{1} \simeq 0.25\,\rm{eV}$ is planned to be achieved.

An important information about the sum of neutrino masses can be  
obtained from cosmological data. From the
Wilkinson Microwave Anisotropy Probe (WMAP)
and 2 degree Field Galaxy Redshift Survey (2dFGRS) data 
it was found \cite{cosmg}
\begin{equation}
\sum_{i}m_{i} \leq 0.7 \,~\rm{eV}.
\label{eq:21}
\end{equation}
More conservative bound was obtained in \cite{teg}
from the analysis of the latest Sloan Digital Sky Survey data and
WMAP data. The best-fit value of 
$\sum_{i}m_{i}$ was found to be equal to zero. For the upper bound 
one obtains
\begin{equation}
\sum_{i}m_{i} \leq 1.7 \,~\rm{eV}.
\label{eq:22}
\end{equation}
For the case of three massive neutrinos this bound implies 
\begin{equation}
m_{1} \leq 0.6\,~\rm{eV}.
\label{eq:23}
\end{equation}

The value of $|m_{\beta \beta}|$ depends on the neutrino
mass spectrum 
\cite{hier2,hier1,hier3,crit,hiera,hier5,hier6,hier4,hier7,hier8}.
We discuss three "standard"  neutrino mass spectra, which are 
frequently considered in the literature:
\begin{enumerate}
\item
{\it 
The normal hierarchy of neutrino masses}\footnote{Notice that masses of charged
leptons, up and down quarks satisfy hierarchy of the type  (\ref{eq:24})
}, which  corresponds to the case
\begin{equation}
m_{1} \ll m_{2} \ll m_{3}. 
\label{eq:24}
\end{equation}
In this case  neutrino masses are known from neutrino oscillation
data.
We have
\begin{equation}
m_{1}\ll 
\sqrt{ \Delta m^{2}_{\rm{sol}}}, ~~~
m_{2}\simeq \sqrt{ \Delta m^{2}_{\rm{sol}}}, ~~~ 
m_{3}\simeq \sqrt{ \Delta m^{2}_{\rm{atm}}}.
\label{eq:25}
\end{equation}
For the effective Majorana mass we have
the following upper and lower bounds
\begin{equation}
|m_{\beta \beta}|\leq \left(\,
\cos^{2} \theta_{13}\, \sin^{2} \theta_{\rm{sol}} \,
 \sqrt{\Delta m^{2}_{\rm{sol}}} + 
\sin^{2} \theta_{13}\, \sqrt{\Delta m^{2}_{\rm{atm}}}\,\right)
\label{eq:26}
\end{equation}
and 
\begin{equation}
|m_{\beta \beta}|\geq\left |\,\cos^{2} \theta_{13}\, \sin^{2}
\theta_{\rm{sol}} \,
\sqrt{\Delta m^{2}_{\rm{sol}}} - 
\sin^{2} \theta_{13}\, \sqrt{\Delta m^{2}_{\rm{atm}}}\,\right|
\label{eq:27}
\end{equation}

Using the best-fit values 
of the solar neutrino oscillation parameters
[see Eqs. (\ref{eq:2}) and (\ref{eq:18})] and the upper
bound  (\ref{eq:19}) we end up with
\begin{eqnarray}
\cos^{2} \theta_{13}\, \sin^{2}
\theta_{\rm{sol}} \,
\sqrt{\Delta m^{2}_{\rm{sol}}} ~~ \simeq~~  2.32\cdot 10^{-3}\rm{eV},
\nonumber\\
\sin^{2} \theta_{13}\,~ \sqrt{\Delta m^{2}_{\rm{atm}}}
~~\leq~~ 2.24\cdot 10^{-3}\rm{eV}.
\label{eq:28}
\end{eqnarray}
We note that the first and the second terms on the
right hand side of Eq. (\ref{eq:27}) 
differ only slightly from each other. It means that the value of
effective Majorana neutrino mass $m_{\beta\beta}$ might be
close to zero. For the choice of three possible values 
$\sin^{2} \theta_{13} =$ 0.05, 0.01 and 0.00 we end up 
with allowed intervals 
for $|m_{\beta\beta}|$:
\begin{eqnarray}
\sin^{2} \theta_{13} &=& 0.05~~ \Rightarrow~~
8.5~10^{-5}~ {eV}~ \leq ~|m_{\beta \beta}|~\leq~ 4.6~10^{-3}~ {eV}, \nonumber\\
 &=& 0.01~~ \Rightarrow~~
2.0~10^{-3}~ {eV}~ \leq ~|m_{\beta \beta}|~\leq~ 2.9~10^{-3}~ {eV}, \nonumber\\
 &=& 0.00~~ \Rightarrow~~|m_{\beta \beta}|~=~2.4~10^{-3}~ {eV}.
\label{eq:29}
\end{eqnarray}
From (\ref{eq:29}) it follows that a smaller value 
of $\sin^{2} \theta_{13}$ implies more narrow
range of the allowed values of $m_{\beta\beta}$.
We also conclude that in the case of the normal neutrino mass hierarchy 
the upper bound $|m_{\beta \beta}|~\leq~ 4.6~10^{-3}~ {eV}$ is far 
from the value which can be reached in the $0\nu\beta\beta$-decay experiments 
of the next generation.

\vspace{0.3cm}

\item {\it Inverted hierarchy of neutrino masses}. It is given 
by the condition
\begin{equation}
m_{3} \ll m_{1} < m_{2}. 
\label{eq:30}
\end{equation}

In this case for neutrino masses we have
\begin{eqnarray}
m_{3} &\ll& 
\sqrt{ \Delta m^{2}_{\rm{atm}}},~~~~
 m_{1}\simeq \sqrt{ \Delta
m^{2}_{\rm{atm}}}, ~\nonumber\\
m_{2} &\simeq& \sqrt{ \Delta m^{2}_{\rm{atm}}}(1 + \frac{\Delta m^{2}_{\rm{sol}}}
{2\,\Delta m^{2}_{\rm{atm}}})\simeq\sqrt{ \Delta m^{2}_{\rm{atm}}}.
\label{eq:31}
\end{eqnarray}
The effective Majorana mass is given by
\begin{equation}
|m_{\beta \beta}|\simeq \sqrt{ \Delta m^{2}_{\rm{atm}}}\,~
|\sum_{i=1,2} \,U_{ei}^{2}\,~|.
\label{eq:32}
\end{equation}
Neglecting small ($\leq 5$\%) corrections due to $|U_{e3}|^2$, 
for $|m_{\beta \beta}|$ we obtain 
\begin{equation}
|m_{\beta \beta}|\simeq \sqrt{ \Delta m^{2}_{\rm{atm}}}\,~
(1-\sin^{2} 2\,\theta_{\rm{sol}}\,\sin^{2}\alpha_{21})^{1/2},
\label{eq:33}
\end{equation}
where $\alpha_{21}=\alpha_{2}-\alpha_{1}$ is Majorana CP-phase
difference.

Thus, in the case of the inverted mass hierarchy 
the value of the effective Majorana mass can lay in the range
\begin{equation}
\cos  2\,\theta_{\rm{sol}}\,\sqrt{ \Delta m^{2}_{\rm{atm}}}\leq
 |m_{\beta \beta}|\leq \sqrt{ \Delta m^{2}_{\rm{atm}}}.
\label{eq:34}
\end{equation}
The bounds in Eq. (\ref{eq:34})
correspond to the case of the CP-conservation:
the upper bound corresponds to the case of the
equal CP parities of 
$\nu_{2}$ and $\nu_{3}$ and the lower bound to the case of the
opposite CP parities.
From (\ref{eq:18}) and (\ref{eq:34}) we get
\begin{equation}
0.42\,~\sqrt{ \Delta m^{2}_{\rm{atm}}}\leq
 |m_{\beta \beta}|\leq \sqrt{ \Delta m^{2}_{\rm{atm}}}.
\label{eq:35}
\end{equation}
Let us assume that the problem of nuclear matrix elements will be solved
(say, in a manner we discussed before). 
If the measured value of $|m_{\beta \beta}|$ will be within the range given in
Eq. (\ref{eq:35}), it will be an indication in favor of inverted hierarchy of neutrino 
masses\footnote{Notice that the type of neutrino mass
spectra (normal or inverted) can be determined via the comparison of 
the probabilities of $\nu_{\mu}\to\nu_{e}$ and
$\bar\nu_{\mu}\to\bar\nu_{e}$ transitions 
in long baseline neutrino experiments \cite{lind}}.
The only unknown parameter, which enter into expression for the
effective Majorana mass in the case of inverted hierarchy, is 
$\sin^{2}\alpha_{21}$. Thus, the measurement of $ |m_{\beta \beta}|$
might allow, in principle, to obtain an information about Majorana CP 
phase difference $|\alpha_{21}|$ \cite{hier1,hier3}. 
It would require, however, a precise measurement of the $0\nu\beta\beta$
half-time.

\vspace{0.3cm}

\item
{\it Almost degenerate neutrino mass spectrum.} 

In the two cases of neutrino mass spectra, discussed above, 
the lightest neutrino mass was assumed to be 
small. The existing bounds on the absolute value of the
neutrino mass (see (\ref{eq:20}) and (\ref{eq:23})) do not exclude the
possibility that the lightest neutrino mass is much larger than
$\sqrt{ \Delta m^{2}_{\rm{atm}}}$. In this case we have
\begin{equation}
m_{1} \simeq m_{2} \simeq  m_{3},
\label{eq:36}
\end{equation}
and the effective Majorana mass take the form
\begin{equation}
|m_{\beta \beta}|\simeq m_{1}\,~
|\sum_{i=1} ^{3}\,U_{ei}^{2}\,~|.
\label{eq:37}
\end{equation}
By neglecting a small contribution of
the parameter $|U_{e3}|^{2}$, from Eq. (\ref{eq:37}) we get
\begin{equation}
|m_{\beta \beta}|\simeq \,m_{1}\,~
(1-\sin^{2} 2\,\theta_{\rm{sol}}\,\sin^{2}\alpha_{21})^{\frac{1}{2}}.
\label{eq:38}
\end{equation}
Using the best fit  value (\ref{eq:18}) we obtain 
\begin{equation}
0.42\,~m_{1}\leq
 |m_{\beta \beta}|\leq m_{1}.
\label{eq:39}
\end{equation}
Thus, if it occurs that the effective Majorana mass 
$|m_{\beta \beta}|$ is relatively large (much larger than 
$\sqrt{ \Delta m^{2}_{\rm{atm}}}\simeq 4.5\cdot 10^{-2}\rm{eV}$) 
it signifies that the neutrino mass spectrum is 
almost degenerate. If the case 
$|m_{\beta \beta}|\gg \sqrt{ \Delta m^{2}_{\rm{atm}}}
\simeq 4.5\cdot 10^{-2}\rm{eV}$ will be confirmed by the
$0\nu\beta\beta$-decay experiments, the explanation could be
a degenerate neutrino mass spectrum.
From (\ref{eq:39}) for the common neutrino mass we get
the range
\begin{equation}
 |m_{\beta \beta}|\leq m_{1}\leq 2.38\,~|m_{\beta \beta}|.
\label{eq:40}
\end{equation}
From (\ref{eq:38}) it is obvious that if the common mass $m_{1}$ 
will be determined from $\beta$-decay measurements and/or 
cosmological data, the evidence of the $0\nu\beta\beta$-decay 
will allow to deduce a valuable information about Majorana CP phase 
difference via the 
accurate measurement of the $0\nu \beta \beta$
half-time.
\end{enumerate}

For the purpose of illustration of the problem of the neutrino mass hierarchy
we will assume that $\ll$ and $\gg$ in (\ref{eq:25}) and (\ref{eq:31}) 
can be represented by a factor 5. Then we have
\begin{eqnarray}
Normal~ hierarchy (NH):&& ~~~~~m_1~ \ll~ \sqrt{\Delta m^{2}_{\rm{sol}}} \nonumber \\
&& ~~~~~m_1 ~\le~ \sqrt{\Delta m^{2}_{\rm{sol}}}/5 = 1.7~10^{-3}~eV,  \nonumber \\       
Inverted~ hierarchy (IH):&& ~~~~~m_3 \ll \sqrt{\Delta m^{2}_{\rm{atm}}} \nonumber \\
&& ~~~~~m_3~ \le~ \sqrt{\Delta m^{2}_{\rm{atm}}}/5 = 8.9~10^{-3}~eV,  \nonumber \\       
Almost~ degenerate (AD):&& ~~~~~m_1, ~m_3~ \gg~ \sqrt{\Delta m^{2}_{\rm{atm}}} \nonumber \\
&& ~~~~~m_1,~ m_3~ \ge~ 5~\sqrt{\Delta m^{2}_{\rm{atm}}} = 0.22~eV. 
\label{eq:41}
\end{eqnarray}
It is worthwhile to notice that 
in the case of the almost degenerate neutrino mass spectrum there is an 
upper limit from the cosmological data [see Eq. (\ref{eq:23})]: 
$m_1,~m_3~\le~0.6~eV$.
The bounds in  (\ref{eq:41}) are displayed in Fig. 1.

\begin{figure}[t]
\epsfxsize=8.0cm
\begin{center}
\epsfig{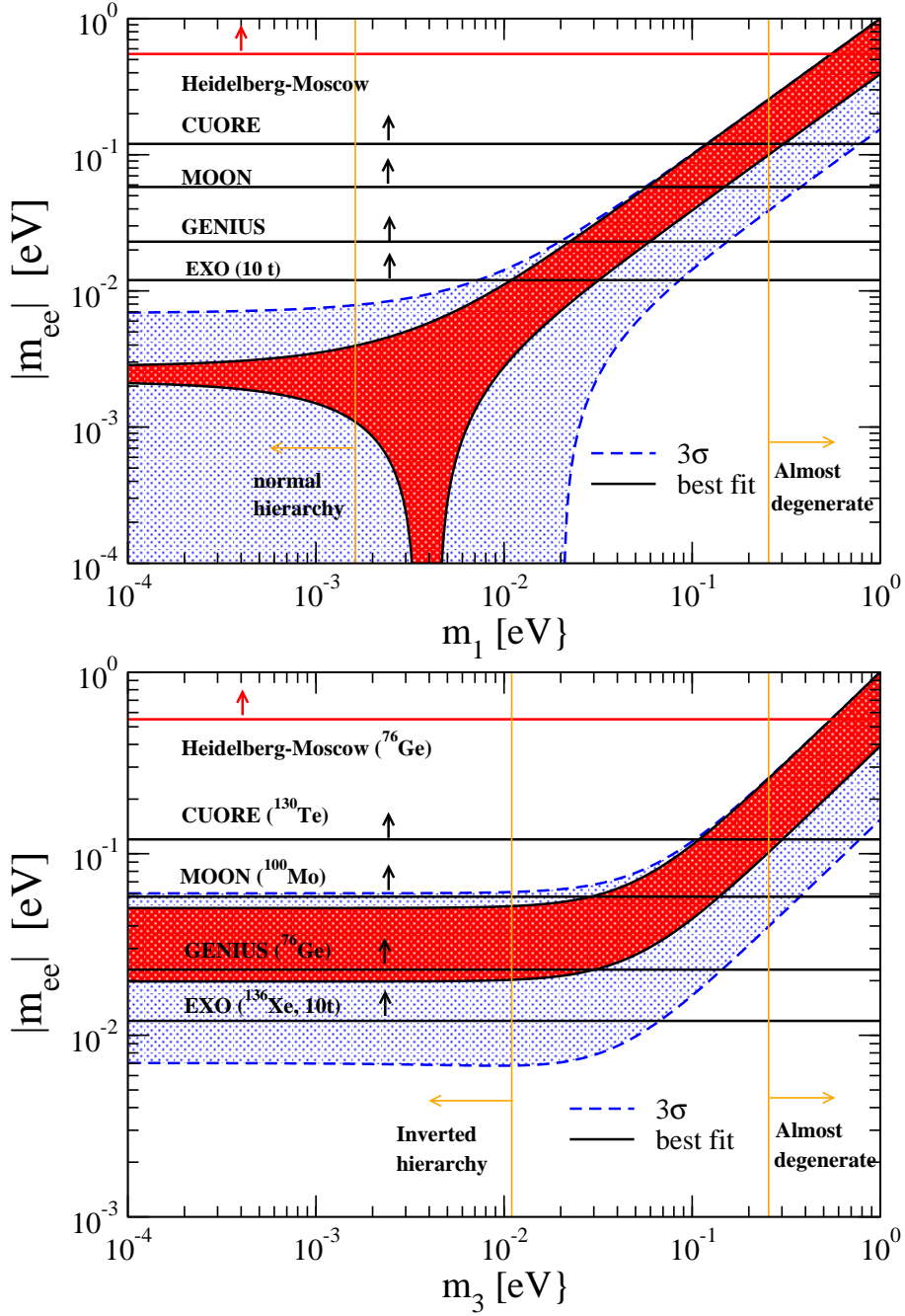}
\end{center}
\caption{The effective Majorana neutrino mass $m_{\beta\beta}$ as function 
of lightest neutrino mass $m_1$
(the normal hierarchy of neutrino masses, the upper panel) and 
$m_3$ (the inverted hierarchy of neutrino masses, the lower panel).
The ranges of the of the normal hierarchy ($m_1 \le 1.7~10^{-3}~eV$)
and inverted hierarchy ($m_3 \le 8.9~10^{-3}~eV$)
of neutrino masses and almost degenerate 
($m_1,~ m_3\ge 0.22~eV$) neutrino mass spectrum [see Eq. (\protect\ref{eq:41}) and
the text above for definition] are indicated by dashed line arrows. 
The best fit results (the region with solid line boundary)
correspond to the parameter set
$\Delta m^{2}_{\rm{sol}} = 7.1~10^{-5} ~{eV}^{2}$,
$\Delta m^{2}_{\rm{atm}}= 2.0~10^{-3}~ {eV}^{2}$ ,
$\sin^{2} \theta_{12}= 0.29$ \protect\cite{SNO,S-K,SNOsalt}
and  $sin^2 \theta_{13} = 0.00$. 
The $3 \sigma$ results (the region with dashed line boundary) 
correspond to the global fit of Ref. 
\protect\cite{malt}. The sensitivities of the future
experiments on the search for the 
$0\nu\beta\beta$-decay of different isotopes are indicated with 
horizontal solid bold lines.
\label{fig.1}}
\end{figure}

In Table \ref{table.2} we give the values of neutrino masses $m_1$, $m_2$
and $m_3$, the minimal and the maximal predicted values of $|m_{\beta\beta}|$ 
in the cases
of the  normal and inverted hierarchy of the neutrino masses
and almost degenerate neutrino mass spectrum.
Three values for $sin^2 \theta_{13}$ compatible with the CHOOZ upper bound 
are considered: 
$sin^2 \theta_{13} =$ 0.00, 0.01 and 0.05. We see that 
by decreasing $sin^2 \theta_{13}$ the allowed interval for
$|m_{\beta\beta}|$ becomes more narrow. 
This behavior is apparent especially
in the case of the normal hierarchy of neutrino masses. For this
scenario of the neutrino mass spectrum the largest value of $|m_{\beta\beta}|$
is of the order of $5~10^{-3}~eV$. None of the planned
$0\nu\beta\beta$-decay experiments can reach such a level of
sensitivity to $|m_{\beta\beta}|$ (see Table \ref{table.1}).
In the case of the inverted hierarchy  $|m_{\beta\beta}|$
depends only weakly on the angle $\theta_{13}$ and its maximal value
is about an order of magnitude larger than in the case of the
of normal hierarchy. This sensitivity
can be reached {without only} by the future Ge $0\nu\beta\beta$-decay experiments
(see Fig. \ref{fig.1}). For this type of neutrino
mass spectrum the maximal and the minimal 
allowed values of $|m_{\beta\beta}|$ differ by about factor 2.5. Thus,
it will be possible to conclude about the Majorana CP 
phase difference in the case of the observation of the 
$0\nu\beta\beta$-decay
with  $|m_{\beta\beta}|$   in the range $1.8-4.5~10^{-2}~eV$, if 
the uncertainties of the nuclear matrix elements are small.
We stress that in order to find some information about CP-phase
difference the value of the lightest neutrino must be known with 
good enough precision. The Tables \ref{table.1}
and \ref{table.2} suggest \footnote{ Let us stress that 
the values of the effective Majorana mass $|m_{\beta\beta}|$, 
given in the Table 1 and Table 2 were obtained
with the nuclear matrix elements of Ref. [21]}
that non-Ge experiments NEMO3, MOON ($^{100}Mo$),
CUORE, ($^{130}Te$), XMASS and  EXO ($^{136}Xe$) will be able
to test mainly the case of the almost degenerate mass spectrum.

\begin{figure}[t]
\epsfxsize=8.0cm
\begin{center}
\epsfig{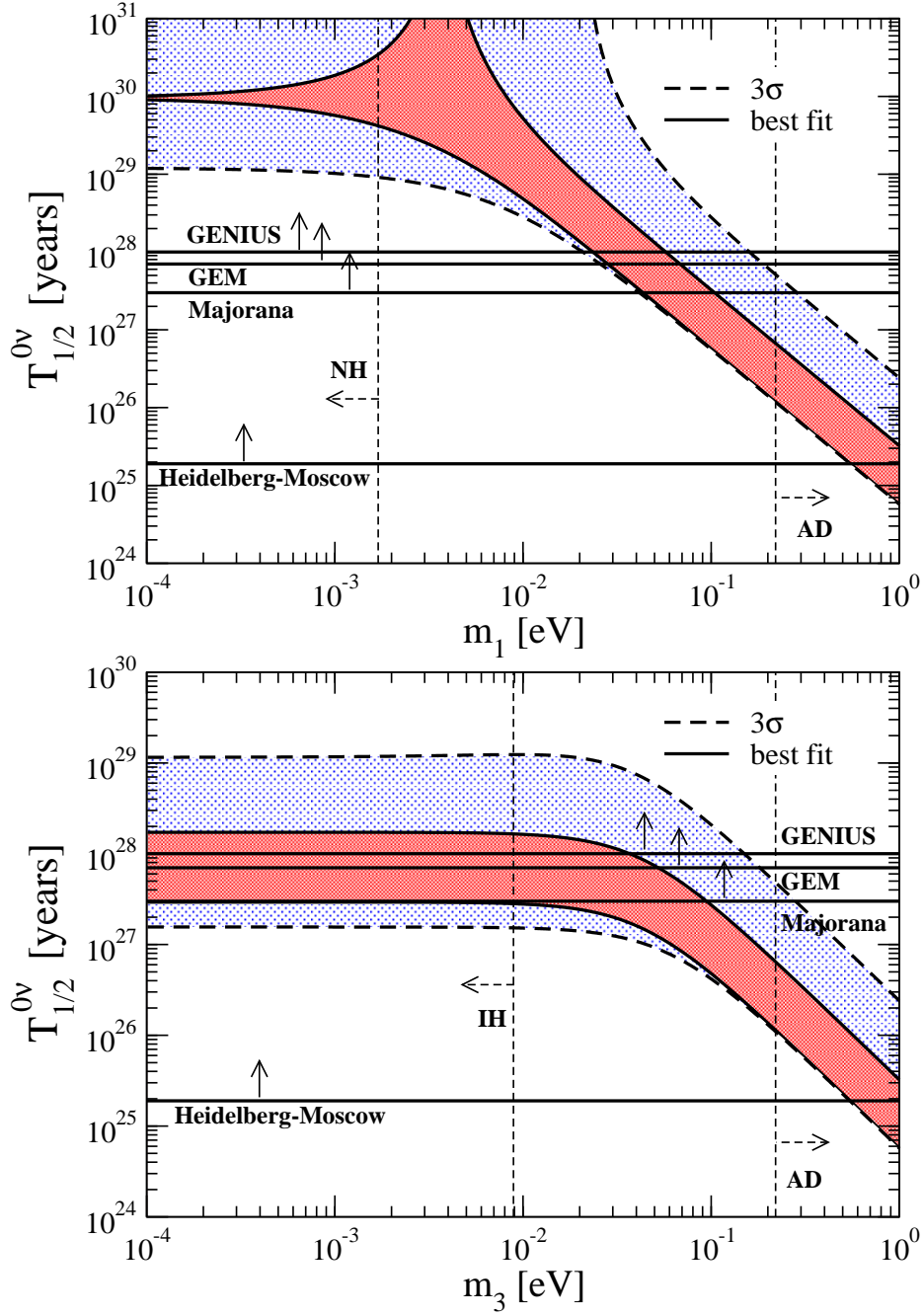}
\end{center}
\caption{The neutrinoless double beta half-life of $^{76}Ge$
as function of the lightest neutrino mass $m_1$ (upper panel)
and $m_3$ (lower panel). Conventions are the same as in 
Fig. \protect\ref{fig.1}. We see that all three planned Ge
experiments Majorana, GEM and GENIUS can check neutrino
mixing scenario of the inverted hierarchy of masses.
NH, IH and AD denote normal hierarchy, inverted hierarchy of
neutrino masses and almost degenerate neutrino mass spectrum,
respectively.
\label{fig.2}}
\end{figure}

\begin{figure}[t]
\epsfxsize=8.0cm
\begin{center}
\epsfig{file=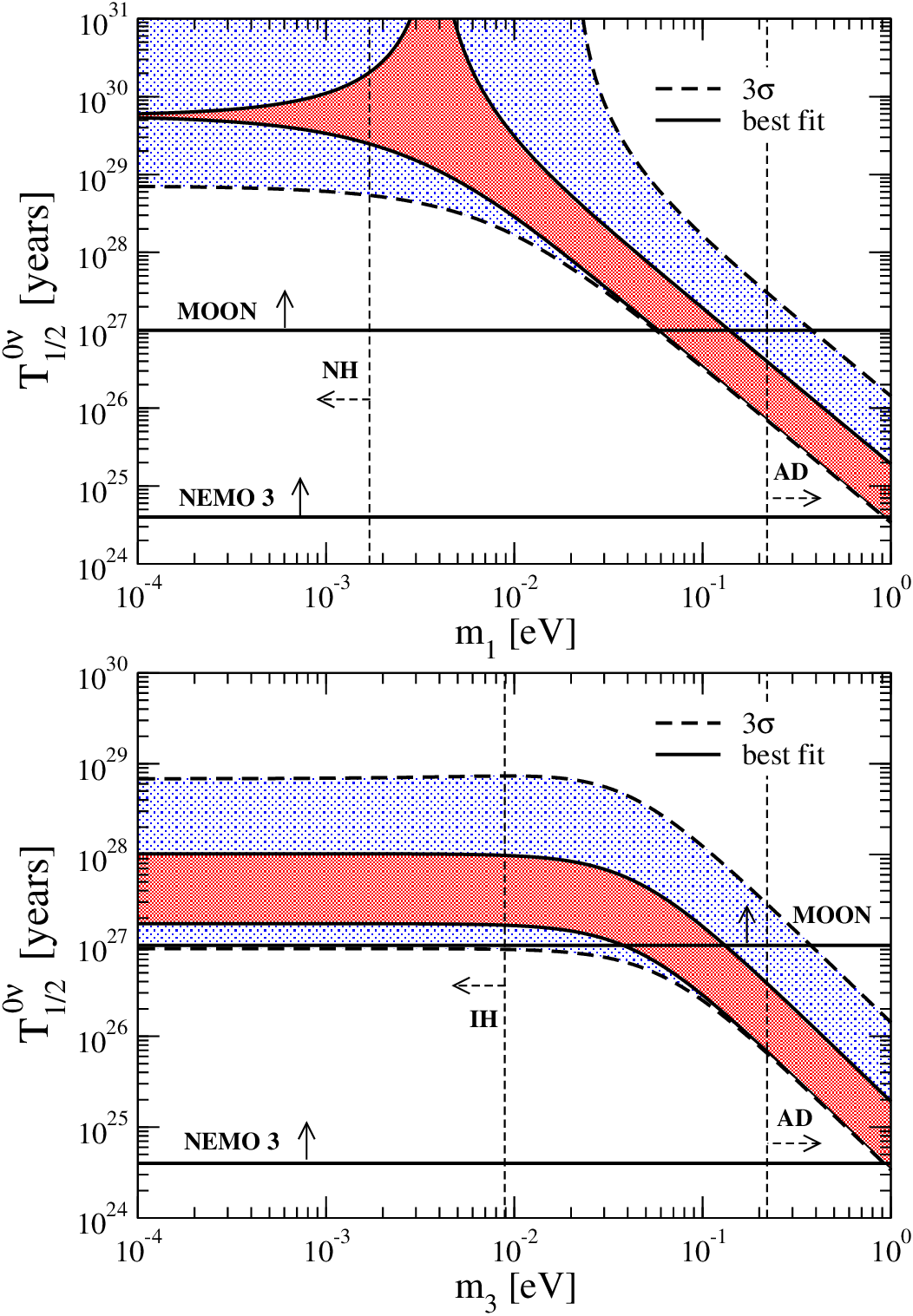,scale=0.69}
\end{center}
\caption{The neutrinoless double beta half-life of $^{100}Mo$
as function of the lightest neutrino mass $m_1$ (upper panel)
and $m_3$ (lower panel). Conventions are the same as in 
Fig. \protect\ref{fig.2}. 
\label{fig.3}}
\end{figure}

\begin{figure}[t]
\epsfxsize=8.0cm
\begin{center}
\epsfig{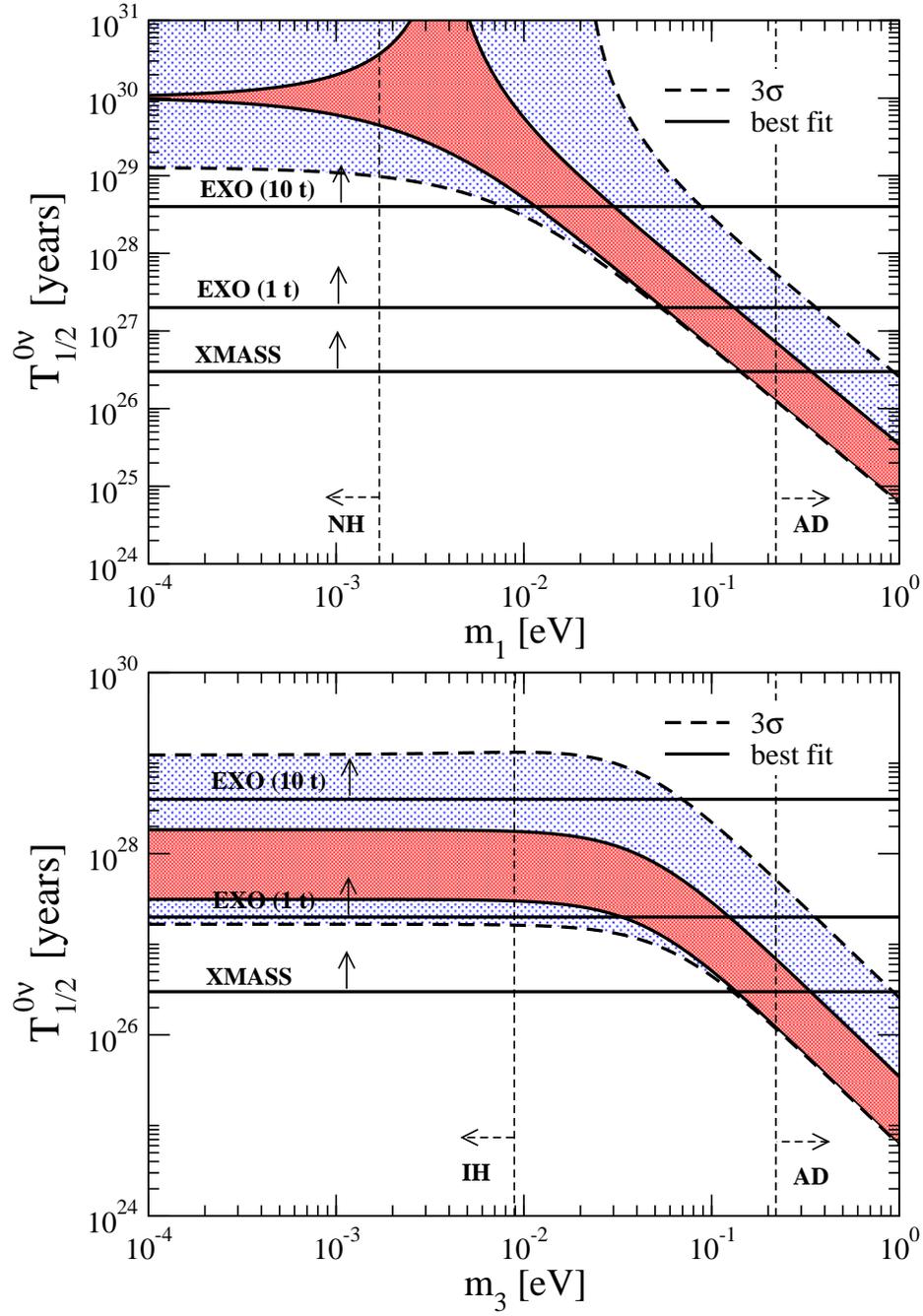}
\end{center}
\caption{The neutrinoless double beta half-life of $^{130}Te$
as function of the lightest neutrino mass $m_1$ (upper panel)
and $m_3$ (lower panel). Conventions are the same as in 
Fig. \protect\ref{fig.2}. 
\label{fig.4}}
\end{figure}

\begin{figure}[t]
\epsfxsize=8.0cm
\begin{center}
\epsfig{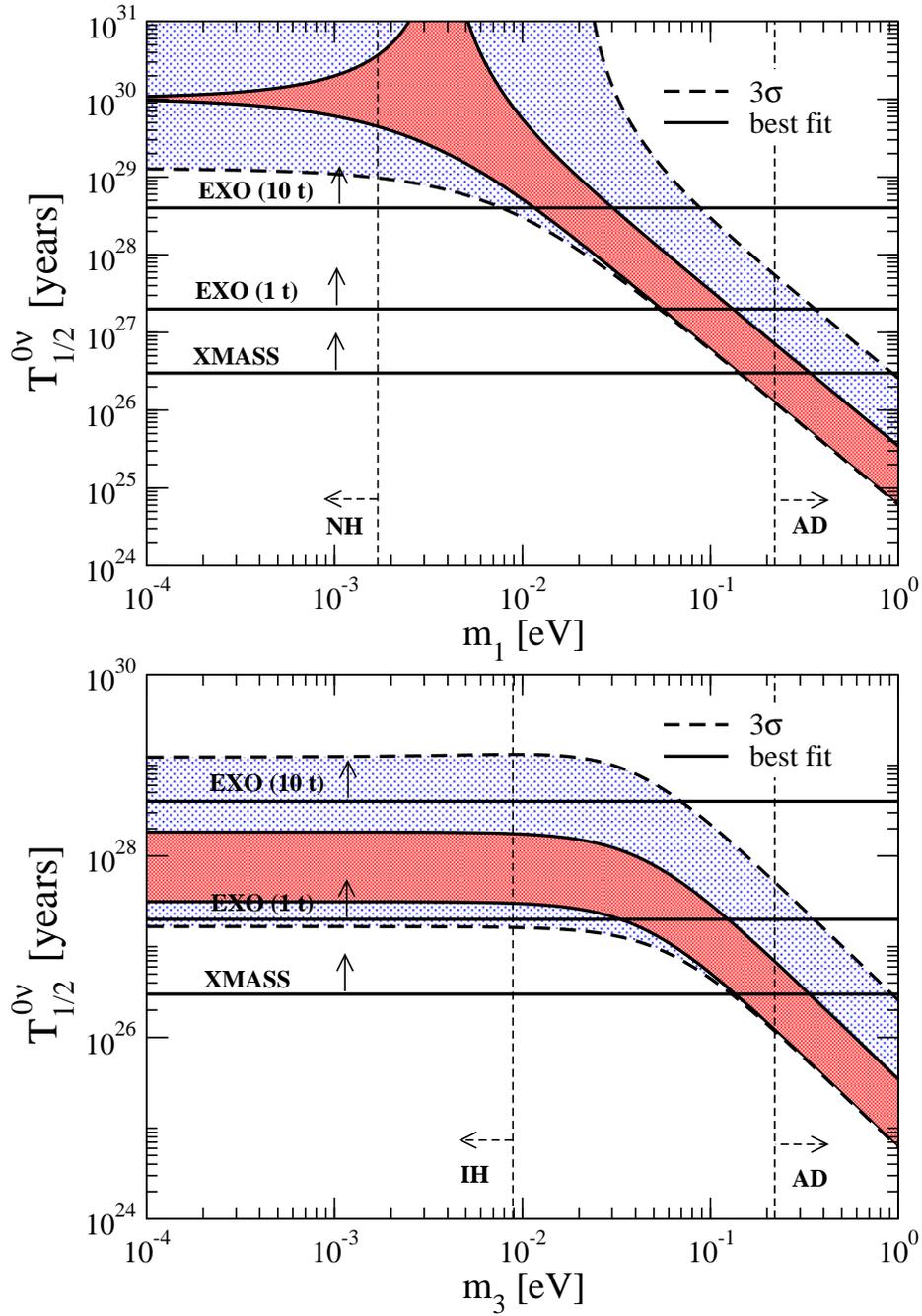}
\end{center}
\caption{The neutrinoless double beta half-life of $^{136}Xe$
as function of the lightest neutrino mass $m_1$ (upper panel)
and $m_3$ (lower panel). We see that the planned EXO (10 t)
experiment can check neutrino
mixing scenario of the inverted hierarchy of masses.
Conventions are the same as in 
Fig. \protect\ref{fig.2}. 
\label{fig.5}}
\end{figure}

A very good potential for discovery of the $0\nu\beta\beta$-decay
have the GEM, GENIUS and Majorana experiments, which plan to use
enriched $^{76}Ge$ source.  In Fig. \ref{fig.2} the
half-life of the $0\nu\beta\beta$-decay of $^{76}Ge$ is plotted 
as function of the lightest neutrino mass. We see that these 
three experiments might observe the $0\nu\beta\beta$-decay
in the case of almost degenerate spectrum and in the case 
of the inverted hierarchy of neutrino masses. Let us stress that
it is very important to achieve high sensitivity   
also in several other experiments
using as a radioactive source other nuclei. It will allow 
to obtain important information about the accuracy
of nuclear matrix elements involved and to discuss the effect
of the CP-Majorana phases. The expected half-lifetimes
of the $0\nu\beta\beta$-decay of $^{100}Mo$, $^{130}Te$ and 
$^{136}Xe$ calculated with nuclear matrix elements of Ref.
\cite{rodin} with minimal neutrino mass considered as a parameter
are shown Figs.  \ref{fig.3}, \ref{fig.4} and \ref{fig.5}).  

\section{Summary and conclusions}

After the discovery of  neutrino oscillations 
in the atmospheric, solar and reactor KamLAND experiments, the problem
of the nature of neutrinos with definite masses (Dirac or Majorana?)
is one of the most important.
The most sensitive process to the possible violation of the lepton number and 
small Majorana neutrino masses is the  neutrinoless double
$\beta$-decay. At present 
many new experiments on the search for the $0\nu\beta\beta$-decay
of $^{76}{Ge}$, $^{100}{Mo}$, $^{130}{Te}$, $^{136}{Xe}$ 
and other nuclei are in preparation or under consideration. 
In these experiments about an order of magnitude improvement of the sensitivity 
to the effective Majorana mass $|m_{\beta \beta}|$ in comparison with 
the current Heidelberg-Moscow \cite{HM} and IGEX \cite{yzd} experiments is expected. 
If the  $0\nu\beta\beta$-decay will be observed 
it will allow not only to establish that massive neutrinos are Majorana particles
but also to reveal the character of the neutrino mass spectrum and 
the absolute scale of neutrino masses.

The data of neutrino oscillation experiments allow to predict 
ranges of possible values of the effective Majorana mass
for different neutrino mass spectra. Thus, in order to discriminate 
different possibilities, it is important not only to observe the  
$0\nu\beta\beta$-decay but also to {\em measure} the effective 
Majorana mass $|m_{\beta \beta}|$. 

From the measured half-lifetime of the $0\nu\beta\beta$-decay
only the product of the effective Majorana mass and the 
nuclear matrix element 
can be determined. There is a wide--spread opinion that the
current uncertainty in the  $0\nu\beta\beta$-decay matrix elements
is of the order of factor three and more \cite{maos}. 
Let us stress that 
a very important source of the uncertainty
is associated with the fixing of the nuclear structure parameter space. 
Recently, surprising results were obtained by fixing of the
particle-particle interaction strength to the $2\nu\beta\beta$-decay
rate \cite{rodin}. This procedure
 allowed to reduce the theoretical uncertainty
of the $0\nu\beta\beta$-decay matrix elements  for
$^{76}{Ge}$, $^{100}{Mo}$, $^{130}{Te}$ and $^{136}{Xe}$
within the QRPA. 
It will be important to confirm this result for other double
beta decaying isotopes and for various QRPA extensions. There is
also a possibility to build a single QRPA theory with all studied 
implementations. The improvement of the calculations of the nuclear 
matrix elements is a real theoretical challenge.
There is a chance that the uncertainty of the
calculated $0\nu\beta\beta$-decay matrix elements will be reduced
down to the order of tenths percent. A possible test of
the calculated nuclear matrix elements will offer an
observation of the $0\nu\beta\beta$-decay of several nuclei.
The spread of the values of $|m_{\beta\beta}|$ associated
with different isotopes will allow to conclude about the quality
of the nuclear structure calculations.

In this paper  we considered  $0\nu\beta\beta$-decay matrix elements
with a reduced theoretical uncertainty \cite{rodin} and
determined the sensitivities of running and planned $0\nu\beta\beta$-decay
experiments to the effective Majorana neutrino mass $|m_{\beta \beta}|$
for of $^{76}{Ge}$, $^{100}{Mo}$, $^{130}{Te}$ and $^{136}{Xe}$. 
The effective Majorana neutrino mass 
was  evaluated also by taking into account
the results of the neutrino oscillation experiments. Three different
cases of neutrino mass spectra were analyzed: i) a normal hierarchical,
ii) an inverted hierarchical, iii) an almost degenerate mass spectrum.
The best fit values for $\Delta m^{2}_{\rm{sol}}$, $\Delta m^{2}_{\rm{atm}}$
and $\sin^{2} \theta_{12}$ were considered . The analysis were performed
for three values of the parameter $\sin^{2} \theta_{13}$
( $\sin^{2} \theta_{13}=~0.00, 0.01, 0.05$). 
A selected group of future experiments associated with the above isotopes
were discussed. It was found that the NEMO3, MOON, CUORE, XMASS and  EXO (1ton)
experiments have chance to confirm or to rule out 
the possibility of the almost degenerate neutrino mass spectrum. 
The planned Ge and Xe experiments (Majorana, GEM, GENIUS and EXO(10t) ) seem to have
a very good sensitivity to $|m_{\beta\beta}|$. These experiments  
will observe the $0\nu\beta\beta$-decay, if the neutrinos are Majorana particles
and there is an inverted hierarchy of neutrino masses.

Finally, by taking into account existing values of neutrino oscillation parameters
we present some  general conclusions:
\begin{itemize}
\item
If the $0\nu\beta\beta$-decay will be not observed in the experiments of
the next generation and 
$$|m_{\beta \beta}|~ \leq~ \rm{~ a~ few}\,~ 10^{-2}~ \rm{eV},$$
in this case either massive neutrinos are Dirac particles or massive 
neutrinos are Majorana particles and normal neutrino mass hierarchy 
is realized in nature. 
The observation of the $0\nu\beta\beta$-decay with 
$$|m_{\beta \beta}|~ \geq~ 4.5~  10 ^{-2}~ \rm{eV}$$
will exclude normal hierarchy of neutrino masses.

\item
If the $0\nu\beta\beta$-decay will be observed and
$$
0.42\,~\sqrt{ \Delta m^{2}_{\rm{atm}}}~\leq~
 |m_{\beta \beta}|~\leq~ \sqrt{ \Delta m^{2}_{\rm{atm}}},$$
it will be an indication in favor of the inverted hierarchy of neutrino masses.

\item
If the $0\nu\beta\beta$-decay will be observed in future experiments 
and 
$$|m_{\beta \beta}|~\gg~ \sqrt{ \Delta m^{2}_{\rm{atm}}},$$
in this case the neutrino mass spectrum is almost degenerate and 
a range for the common neutrino mass can be determined.

\item
If from the future tritium neutrino experiments or from future cosmological
measurements the common neutrino mass will be determined, 
it will be possible to predict the value of effective Majorana 
neutrino mass:
$$0.42~m_1\leq |m_{\beta \beta}|\leq  m_1.$$
A non-observation of the $0\nu\beta\beta$-decay with effective
Majorana mass $|m_{\beta \beta}|$ in this range will mean that 
neutrinos are Dirac particles (or other mechanisms of the violation 
of the lepton number are involved). 

\end {itemize}

{\bf Acknowledgments}

This work was supported by the 
DFG under contracts FA67/25-3, 436 SLK 113/8 and GRK683, 
by the State of Baden-W\"{u}rt\-tem\-berg, LFSP "Low Energy Neutrinos", 
by the VEGA Grant agency of the Slovac Republic under contract No.~1/0249/03.  
S. B. acknowledges the support of  the Alexander von Humboldt Foundation and 
the Italien Program ``Rientro dei Cervelli''.

\begin{table}
\caption{The current upper limits on  effective Majorana 
neutrino mass $|m_{\beta\beta}|$ and the sensitivities of  
the future $0\nu\beta\beta$-decay experiments to this parameter 
for A=76, 100, 130 and 136 nuclei. The 
$0\nu\beta\beta$-decay matrix elements $M^{0\nu}$
with reduced uncertainty  \protect\cite{rodin} were used.
In their calculation the $2\nu\beta\beta$-decay matrix element 
$M^{2\,\nu-exp}_{GT}$ deduced from the half-life $T^{2\nu-exp}_{1/2}$
were considered. ${\cal R}^{2\nu/0\nu}$ is
the ratio of the  $2\nu\beta\beta$-decay and   $0\nu\beta\beta$-decay
matrix elements (see Eq. (\protect\ref{eq:9})).
  $T^{0\nu}_{1/2}$ denotes the current lower
limit on the $0\nu\beta\beta$-decay half-life or the
sensitivity of planned $0\nu\beta\beta$-decay experiments.
The symbols
$^*$ and $^\dagger$ indicate the future sensitivity to $|m_{\beta\beta}|$ 
of already running and the planned $0\nu\beta\beta$-decay experiments, 
respectively. HM denotes Heidelberg-Moscow experiment.
\label{table.1}}
\begin{tabular}{|lccccccc|}\hline\hline 
nucl. & $M^{0\nu}$ & $M^{2\,\nu-exp}_{GT}$ &
${\cal R}^{2\nu/0\nu}$ & $T^{2\nu-exp}_{1/2}$ Ref. 
& $T^{0\nu}_{1/2}$ Ref.  & Exp. & $|m_{\beta\beta}|$ \\ \cline{2-8}
 & & $MeV^{-1}$ & $MeV^{-1}$ & years & years & & eV \\ \hline 
$^{76}Ge$ & 2.40 & 0.15 & 0.063 & $1.3~ 10^{21}$\cite{vog02} &
$1.9~ 10^{25}$\cite{HM}  & HM & 0.55 \\
 & & & & &
 $3 ~ 10^{27}$\cite{vog02}  & Majorana & $0.044^\dagger$ \\
 & & & & &
 $7 ~ 10^{27}$\cite{vog02}  & GEM & $0.028^\dagger$ \\
 & & & & &
 $1 ~ 10^{28}$\cite{vog02}  & GENIUS & $0.023^\dagger$ \\
$^{100}Mo$ & 1.16 & 0.22 & 0.19 & $8.0~ 10^{18}$\cite{vog02} &
$6.0~ 10^{22}$\cite{nemo}  & NEMO3 & 7.8 \\
 & & & & &
 $4 ~ 10^{24}$\cite{vog02}  & NEMO3 & $0.92^{*}$ \\
 & & & & &
 $1 ~ 10^{27}$\cite{vog02}  & MOON & $0.058^\dagger$ \\
$^{130}Te$ & 1.50 & 0.017 & 0.013 & $6.1~ 10^{20}$\cite{cunew} &
$1.4~ 10^{23}$\cite{cunew}  & CUORE & 3.9 \\
 & & & & &
 $2 ~ 10^{26}$\cite{vog02}  & CUORE & $0.10^*$ \\
$^{136}Xe$ & 0.98 & 0.030 & 0.031 & $\ge 8.1~ 10^{20}$\cite{vog02} &
$1.2~ 10^{24}$\cite{dama}  & DAMA & 2.3 \\
 & & & & &
 $3 ~ 10^{26}$\cite{vog02}  & XMASS & $0.10^\dagger$ \\
& & & & &
 $2 ~ 10^{27}$\cite{gratta}  & EXO (1t) & $0.055^\dagger$ \\
& & & & &
 $4 ~ 10^{28}$\cite{gratta}  & EXO (10t) & $0.012^\dagger$ \\
\hline\hline 
\end{tabular}
\end{table}

\newpage

\begin{table}
\caption{The effective Majorana neutrino mass $|m_{\beta\beta}|$
in the cases of the normal and inverted hierarchy of neutrino 
masses and the
almost degenerate neutrino mass spectrum [see Eq. (\protect\ref{eq:41}) and
the text above]. The best fit values 
$\Delta m^{2}_{\rm{sol}} = 7.1~10^{-5} ~{eV}^{2}$,
$\Delta m^{2}_{\rm{atm}}= 2.0~10^{-3}~ {eV}^{2}$ and 
$\sin^{2} \theta_{12}= 0.29$ are considered \protect\cite{SNO,S-K,SNOsalt}.
The results are presented for three values of  angle
$\theta_{13}$ from the CHOOZ allowed range
 $sin^2 \theta_{13} \le 0.05$ \protect\cite{fogli}.
\label{table.2}}
\begin{tabular}{|ccccc|}\hline\hline 
 & \multicolumn{3}{c}{ 
Normal hierarchy of $\nu$ masses: $m_{1} \ll m_{2} \ll m_{3}$} &  \\
  $m_1$ [$10^{-3}$ eV] &    $m_2$ [$10^{-3}$ eV] &  $m_3$ [$10^{-2}$ eV]
 &  $sin^2 \theta_{13}$ &
$|m_{\beta\beta}|$ [$10^{-3}$ eV]  \\ \hline
 $(0,~1.7)$ &  $(8.43,~8.60)$ &  
$(4.47,~4.48)$ & 0.00 & 
$(1.29,~3.70)$   \\ 
 & & & 0.01 & $(0.83,~4.11)$  \\ 
 & & & 0.05 & $(0.00,~5.75)$  \\ 
\hline \hline 
 & \multicolumn{3}{c}{ 
Inverted hierarchy of $\nu$ masses: $m_{3} \ll m_{1} < m_{2}$} & \\
 $m_3$ [$10^{-3}$ eV] &    $m_1$ [$10^{-2}$ eV] &  $m_2$ [$10^{-2}$ eV]
 &  $sin^2 \theta_{13}$ &
$|m_{\beta\beta}|$ [$10^{-2}$ eV]  \\ \hline
  $(0,~8.9)$ &  $(4.39,~4.48)$ &  
$(4.47,~4.56)$ & 0.00 & 
$(1.82,~4.50)$  \\ 
  & & & 0.01 & $(1.80,~4.47)$  \\ 
  & & & 0.05 & $(1.72,~4.32)$  \\ 
\hline \hline 
 & \multicolumn{3}{c}{ 
  Almost degenerate $\nu$ mass spectrum: $m_{1} \simeq m_{2} \simeq  m_{3}$} & \\
 $m_1$ [eV] &    $m_2$ [eV] &  $m_3$ [eV]
 &  $sin^2 \theta_{13}$ &
$|m_{\beta\beta}|$ [eV]  \\ \hline
  $(0.22,~0.60)$ &  $(0.22,~0.60)$ &  
$(0.22,~0.60)$ & 0.00 & 
$(0.092,~0.60)$  \\ 
  & & & 0.01 & $(0.089,~0.60)$  \\ 
  & & & 0.05 & $(0.077,~0.60)$  \\ 
\hline\hline 
\end{tabular}
\end{table}

\newpage

\end{document}